\documentclass[twocolumn]{aastex631}
\usepackage{CLASSY_Preamble}

\begin{document}
\submitjournal{AASJournal ApJ}
\shortauthors{Arellano-C\'ordova et al.}
\shorttitle{Rest-frame optical spectra of high-$z$ galaxies}
\title{A First Look at the Abundance Pattern - O/H, C/O, and Ne/O - in $z>7$ Galaxies with JWST/NIRSpec}

\author[0000-0002-2644-3518]{Karla Z. Arellano-C\'{o}rdova}
\affiliation{Department of Astronomy, The University of Texas at Austin, 2515 Speedway, Stop C1400, Austin, TX 78712, USA}

\author[0000-0002-4153-053X]{Danielle A. Berg}
\affiliation{Department of Astronomy, The University of Texas at Austin, 2515 Speedway, Stop C1400, Austin, TX 78712, USA}

\author[0000-0002-0302-2577]{John Chisholm}
\affiliation{Department of Astronomy, The University of Texas at Austin, 2515 Speedway, Stop C1400, Austin, TX 78712, USA}

\author[0000-0002-7959-8783]{Pablo Arrabal Haro}
\affiliation{NSF’s National Optical-Infrared Astronomy Research Laboratory, 950 N. Cherry Ave., Tucson, AZ 85719, USA}

\author[0000-0001-5414-5131]{Mark Dickinson}
\affiliation{NSF’s National Optical-Infrared Astronomy Research Laboratory, 950 N. Cherry Ave., Tucson, AZ 85719, USA}

\author[0000-0001-8519-1130]{Steven L. Finkelstein}
\affiliation{Department of Astronomy, The University of Texas at Austin, 2515 Speedway, Stop C1400, Austin, TX 78712, USA}

\author[0000-0002-6085-5073]{Floriane Leclercq}
\affiliation{Department of Astronomy, The University of Texas at Austin, 2515 Speedway, Stop C1400, Austin, TX 78712, USA}

\author[0000-0002-0361-8223]{Noah S. J. Rogers}
\affiliation{Minnesota Institute for Astrophysics, University of Minnesota, 116 Church Street SE, Minneapolis, MN 55455, USA}

\author[0000-0002-6386-7299]{Raymond C. Simons}
\affiliation{Space Telescope Science Institute, 3700 San Martin Drive, Baltimore, MD, 21218 USA}

\author[0000-0003-0605-8732]{Evan D. Skillman}
\affiliation{Minnesota Institute for Astrophysics, University of Minnesota, 116 Church Street SE, Minneapolis, MN 55455, USA}

\author[0000-0002-1410-0470]{Jonathan R. Trump}
\affiliation{Department of Physics, 196 Auditorium Road, Unit 3046, University of Connecticut, Storrs, CT 06269}

\author[0000-0001-9187-3605]{Jeyhan S. Kartaltepe}
\affiliation{Laboratory for Multiwavelength Astrophysics, School of Physics and Astronomy, Rochester Institute of Technology, 84 Lomb Memorial Drive, Rochester, NY
14623, USA}

\correspondingauthor{Karla Z. Arellano-C\'ordova} 
\email{kzarellano@austin.utexas.edu}


\begin{abstract}
We analyze the rest-frame near-UV and optical nebular spectra of three $z > 7$ 
galaxies from the Early Release Observations taken with the Near-Infrared Spectrograph 
(NIRSpec) on the James Webb Space Telescope (JWST). 
These three high-$z$ galaxies show the detection of several strong-emission nebular lines,
including the temperature-sensitive [\ion{O}{3}] \W4363 line, 
allowing us to {\it directly} determine the nebular conditions and 
abundances for O/H, C/O, and Ne/O.
We derive O/H abundances and ionization parameters that are generally consistent
with other recent analyses. 
We analyze the mass-metallicity relationship (i.e., slope) and its redshift evolution by comparing between the three $z > 7$ galaxies and local star-forming galaxies.
We also detect the \ion{C}{3}] \W\W1907,1909 emission in a $z > 8$ galaxy
from which we determine the most distant C/O abundance to date.
This valuable detection of log(C/O) = $-0.83\pm0.38$ provides the first test of C/O redshift evolution out to high-redshift. 
For neon, we use the high-ionization [\ion{Ne}{3}] \W3869 line to measure the first 
Ne/O abundances at $z>7$, finding no evolution in this $\alpha$-element ratio.
We explore the tentative detection of [\ion{Fe}{2}] 
and [\ion{Fe}{3}] lines in a $z>8$ galaxy, which would indicate a rapid build up of metals. 
Importantly, we demonstrate that properly flux-calibrated and higher S/N spectra are crucial 
to robustly determine the abundance pattern in $z>7$ galaxies with NIRSpec/JWST.

\end{abstract} 


\section{Introduction}

Theory predicts that the first galaxies were composed nearly entirely of hydrogen 
and helium with trace amounts of heavier metals. 
As massive stars synthesized metals in their cores and injected these metals into the 
interstellar medium (ISM), galaxies quickly built up their metal content over time \citep{tinsley80}. 
This chemical enrichment of galaxies has been observed at moderate and low-redshifts with the establishment and evolution of the mass-metallicity relation \citep[MZR; e.g.,][]{tremonti04,dalcanton07,peeples11,zahid14}. 
However, the compositions of the first galaxies, and the ancestors of local galaxies, have remained beyond characterization due to the faint nature and redshifted emission lines. 
The ISM of such early universe galaxies, where little chemical evolution has occurred,  
provides benchmarks for our understanding of how galaxies form and evolve
via their chemical enrichment pathways.

Additionally, there is a general consensus that metal-poor, low-mass galaxies 
hosted a large fraction of the star formation in the high-redshift Universe and 
were, therefore, the dominant contributors to reionizing the early universe
\citep[e.g.,][]{wise14,madau15}.
Determining the precise epoch and source of reionization is currently a major focus of observational cosmology, yet, the relative ionizing radiation contributions from 
stellar versus accretion activity are still uncertain \citep[e.g.,][]{fontanot14}. 
Depending on the model assumptions
(i.e., redshift, luminosity function, intergalactic medium [IGM] clumping factor, 
and ionizing photon production efficiency) 
the estimated escape fraction of ionizing radiation needed to sustain cosmic 
reionization ranges from $< 5$\% \citep{finkelstein19} up to 21\% \citep{naidu20}. 
On the other hand, recent observations have emphasized the role of dust in 
regulating the escape of ionizing photons based on the direct empirical relation 
between the observed Lyman continuum (LyC) escape fraction and the slope of the 
far-ultraviolet (FUV) stellar continuum \citep[e.g.,][]{saldana-lopez22, chisholm22}.
Given the strong trend between dust content and metallicity \citep[e.g.,][]{li19,shapley20},
accurate abundance determinations are crucial to understanding the escape of ionizing photons from galaxies within the epoch of reionization.
Further, recent $z \sim 3$ observations highlight the impact of evolving  abundance ratios
on the production and hardness of ionizing photons \citep{steidel16}.  
The first galaxies shaped the conditions within the IGM and set the initial conditions 
for all subsequent galaxy evolution.  
A detailed study of chemically unevolved galaxies at early epochs is required to establish
the physical conditions that produced the ionizing photons to reionize the early universe.

Very few high-redshift galaxies (z $>$ 3) have been observed in spectroscopic
detail owing to the challenges of observing their rest-frame optical spectra 
through the ground-based transparency windows of the infrared.
Fortunately, the successful launch of the James Webb Space Telescope (JWST)
has opened a new window on high-redshift galaxy spectroscopy, 
pushing the FUV and optical spectroscopic frontiers to higher redshifts than ever before.
However, the success of such observations hinges on our ability to interpret the
gaseous conditions, especially the abundance pattern, that power the observed spectral features. 

Here we examine the Early Release Observations (EROs) of the Near Infrared Spectrograph \citep[NIRSpec,][]{jakobsen22} on JWST of the three $z>7$ galaxies 
in the SMACS J0723.3-732 cluster field and interpret their nebular properties.
We specifically focus on our ability to determine robust nebular properties, such as the electron temperature and ionization parameter, and the O/H, C/O, Ne/O, and Fe/O abundance composition.
While O/H is the standard measure of a galaxies metallicity, the relative abundances of non-$\alpha$ elements (e.g., C, N, and Fe) have been observed to significantly deviate from solar, especially in the metal-poor environments \citep[e.g.,][]{berg19,berg21a}.\looseness=-2

Relative C and Fe abundances are especially important, 
as they are significant contributors to interstellar dust \citep[e.g.,][]{draine11},  
tracers of the physical conditions in nebular gas, and 
important sources of opacity in stars, affecting the time evolution of stellar winds and isochrones.
Additionally, enhanced $\alpha$/Fe ($\alpha$/C) abundance ratios 
could produce harder radiation fields than expected from the O/H abundances in moderate-metallicity (low-metallicity; 
$Z_{neb}. < 0.1 Z_\odot$) galaxies 
\citep[e.g.,][]{steidel18, shapley19, topping20}. 
Additionally, C is observable in high-redshift galaxies both in the rest-frame ultraviolet via \ion{C}{3}] \W1907,[\ion{C}{3}] \W1909 \citep[e.g.,][]{stark16,hutchison19} and the rest-frame far-infrared via \ion{C}{2} 158$\mu$m \citep[e.g.,][]{lagache18}, providing two paths to estimate the C abundance.
Because C is primarily produced in lower mass stars than O, 
the injection of C and O to the interstellar medium (ISM) occurs on 
different time scales, providing a probe of the duration, history, and 
burstiness of the star formation \citep[e.g.,][]{henry00,berg19}.

The remainder of this paper is structured as follows:
In Section \ref{sec:2} we describe the NIRSpec observations and data analysis. Section~\ref{sec:3} describes the measurements of the redshift and the nebular 
emission lines. 
In Section~\ref{sec:4}, we discuss the calculations of nebular abundances,
focusing on the oxygen abundance, the C/O abundance ratio, and the Fe/O flux ratio.
Section~\ref{sec:discussion} explores the O/H, C/O, and Ne/O evolution of galaxies from $z\sim0 - 8.5$.
Our final remarks are summarized in Section~\ref{sec:conclusion}.
Throughout this paper we adopt a flat FRW metric with $\Omega_{\mbox{m}}$ = 0.3, 
$\Omega_{\Lambda}$ = 0.7, and H$_0$ = 70 km s$^{-1}$ Mpc$^{-1}$, and the
solar metallicity scale of \citet{asplund21}, where 
12 + log(O/H)$_\odot = 8.69$, 
log(C/O)$_\odot = -0.23$,
log(Ne/O)$_\odot = -0.63$, and
log(Fe/O)$_\odot = -1.23$.

\section{Sample Properties and Observations}\label{sec:2}
We consider here the three lensed galaxies of the SMACS J0723.3-7327 galaxy cluster at $z>7$, which have the JWST source IDs of 
s04590, s06355, and s10612.
The object ID numbers come from the input source catalog used for the NIRSpec micro-shutter array \citep[MSA;][]{ferruit22} configuration planning.
These objects are gravitationally-lensed galaxies by a massive galaxy cluster at $z = 0.3877$, discovered by the southern extension of the Massive Cluster Survey \citep[MACS;][]{ebeling01, repp18}. 

This cluster was observed with NIRSpec of $\JWST$ as a part of the Early Release Observations released on July 12, 2022 (program ID: 2736; \citealt{pontoppidan:22}).
NIRSpec medium resolution gratings cover a wavelength range of 0.6$\mu$m to 5.2$\mu$m at spectral resolution  $R = \lambda/\Delta \lambda = 1000$ \citep{jakobsen22}. For these observations each 1D spectrum was produced from the combination of two gratings/filters, G235M/F170LP (1.75-3.15 $\mu$m) and G395M/F290LP (2.9-5.2 $\mu$m) with a total exposure time of 8754 seconds in each of the two visits. 

For the computation and analysis of the physical properties and chemical abundances 
of these high$-z$ galaxies, we use the post-processed NIRSpec spectra presented in 
\citet{trump22}, which uses the pipeline of \citet{arrabalharo22}, and compare it to
the original JWST Science Calibration Pipeline v1.6.1\footnote{\url{https://jwst-pipeline.readthedocs.io/en/latest/jwst/introduction.html}} data. 
In order to avoid emission from other sources or detector artifacts,
the 1D post-processed spectra were extracted using a narrow custom aperture centered at 
the location of the emission peaks within each slit. 
The post-processing flux calibration was derived from the response of flat-spectrum 
point-like sources in the Cosmic Evolution Early Release Science 
\citep[CEERS, PI: S. Finkelstein,][]{finkelstein22a} simulations performed making use 
of the NIRSpec Instrument Performance Simulator \citep{piqueras10}. 
The details of this new post-processed data will be presented in \citet{arrabalharo22}.

The resulting post-processed spectra of s04590, s06355, and s10612 are shown in 
Figure~\ref{fig:spectra}.
The two visits of both s06355 and s10612, o007 and o008, are very similar in both
flux and noise levels.
However, even after the post-processing, we find the same emission-line flux 
measurements offsets for the two visits of s04590 as \citet{trump22}.
As a result, these authors recommend only using emission-line ratios of nearby lines.
Further, \citet{curti22a} recently inspected the 2D NIRspec spectra and 
revealed an issue with the spectrum of the o007 visit of s04590 because the 
shutter[3,27,167] did not open. 
Therefore, we separate and call out calculations of the metallicity and other physical parameters of s04590 made with the o007 observation for this reason. 
However, we use the o007 observations only for comparison reasons in 
Sec.~\ref{sec:artifact}.
This allows us to have consistent estimates of the physical properties of the 
gas despite the caveats due to absolute flux calibration of the data \citep[see discussion in][]{trump22}. 


\section{Measurements}
\label{sec:3}
The JWST/NIRspec spectra detect several rest-frame optical nebular emission lines in the two different visits (o007 and o008) for s04590, s06355, and s10612, which provide the opportunity to 
measure the redshift and important physical properties, including the reddening, 
the total and relative metallicities, and the ionization parameter. 
 
\subsection{Source Redshift}
We have measured the systemic redshift of s04590,s06355, and s10612 using the most 
reliable strong emission lines in the spectra covered by NIRSpec. 
Specifically, we measured the observed central wavelengths of the five strongest emission 
lines in the NIRSpec spectrum (H$\delta$, H$\gamma$, H$\beta$, and [\ion{O}{3}] 
\W\W4959,5007) using Gaussian profile fits.
Relative to the rest-frame vacuum wavelengths, we determine average redshifts of 
8.495 for s04590, 
7.665 for s06355, and 
7.659 for s10612. 
Our estimated redshifts are in agreement with those published recently by \citet{carnall22}, \citet{schaerer22}, \citet{trump22} and \citet{brinchmann22}.

\subsection{Nebular emission lines}
Nebular emission lines are an important source of information, providing the 
chemical composition and physical properties of the ionized gas 
\citep[e.g.,][]{kewley+19, maiolino+19}.  
In this context, significant rest-frame UV and optical emission line detections 
are essential to analyzing the physical properties of galaxies across cosmic time. 
Therefore, the accurate emission-line ratios are vital to determining extragalactic 
metallicities and, subsequently, constraining fundamental galaxy trends, such as the 
mass-metallicity relationship across cosmic time.

To ensure accurate measurements of the emission lines in the NIRspec observations,
we use IRAF\footnote{IRAF is distributed by the National Optical Astronomy Observatory, which is operated by the Association of Universities for Research in Astronomy, Inc., under cooperative agreement with the National Science Foundation.} 
to carefully inspect and measure the fluxes of each individual nebular emission 
feature in each of the two visits available per galaxy.
Specifically, we measured the fluxes by first determining the local continuum on 
each side of an isolated emission line and then integrating above the continuum. 
For those lines that appear blended, we fit multi-component Gaussian profiles. 
The errors in the flux were calculated using the expression reported in 
\citet{berg13} and \citet{ Rogers21}, which accounts for uncertainties in the flux calibration and 
sky subtraction.
For the NIRspec spectra, the dominant uncertainty comes from the flux calibration.
Similar to \citet{trump22}, we find a factor two of difference between the absolute
fluxes of the o007 and o008 visits of s04590, while the two visits of
s06355 and s10612 match very well.
To account for the potentially very large uncertainty in absolute flux calibration,
we increased the flux calibration uncertainty to 20\%\ and \citep[following the recommendation of][]{trump22} use nearby line ratios when possible.  
In Table~\ref{tab:intensities}, we reported the observed and corrected intensities with respect to H$\beta$ for the three high-$z$ galaxies. Such optical emission lines are detected with a signal-to-noise (S/N) or $F_\lambda$/$\delta F\lambda$ larger than 3.
The UV lines reported in Table~\ref{tab:intensities} for s04590 show detection with a  S/N < 3 (i.e., \ion{C}{3}]~\W1907, 1909).

\subsection{Reddening}
The large uncertainty associated with the flux calibration provides a 
particular challenge for determining the reddening due to dust from the
Balmer decrement, which spans a large wavelength range.
We, therefore, determine a primitive estimate of the reddening due to dust 
using the differences between the observed and theoretical values of the 
Balmer decrement{\b, and by
assuming Case B recombination and typical nebular conditions}
(i.e., $T_ e = 18,000$ K, and $n_e = 100$ cm$^{-3}$) 
corresponding to theoretical values of H$\gamma$/H$\beta$ = 0.47 and H$\delta$/H$\beta$ = 0.26 (calculated using the \texttt{PyNeb} package \citep{luridiana15}). 

For all three galaxies, the H$\gamma$/H$\beta$ ratios are larger than physical
but consistent between the two different visits (o007 and o008) in each galaxy. The observed values of H$\gamma$/H$\beta$ range between 0.40--0.50, which implies  $E(B-V)$ values between $-0.09$ and $-$0.32. In particular, the spectrum of the o007 visit of s04590 provide an observed value of  H$\gamma$/H$\beta$ $\sim$ 1 \citep[see also discussion in][]{trump22}.
We, therefore, calculate the color-excess, $E(B-V)$, based on only 
H$\delta$/H$\beta$ (0.20--0.22). 
Note that we used the $E(B-V)$ value derived from the o008 visit of
s04590 to correct the o007 spectrum. 
Emission line fluxes were then corrected for reddening using the \citet{cardelli89} 
reddening law for the rest-frame optical lines and the \citet{reddy16} reddening law
for the rest-frame UV lines.

The resulting $E(B-V)$ values are reported in Table~\ref{tab:metal}, with values 
range between 0.31--0.34.
Such $E(B-V)$ values are typical of $z\sim0$ galaxies, but are perhaps surprising
for such chemically-young high-$z$ galaxies. 
Given the significant flux calibration uncertainty, we adopt these $E(B-V)$ values
as upper limits and explore their effect by determining abundances assuming both 
the $E(B-V)$ upper limits and a $E(B-V)$ lower limit of 0 in Section~\ref{sec:O/H}.

\begin{figure*}
\begin{center}
    \includegraphics[width=0.90\textwidth, trim=35mm 5mm 35mm 5, clip=yes]{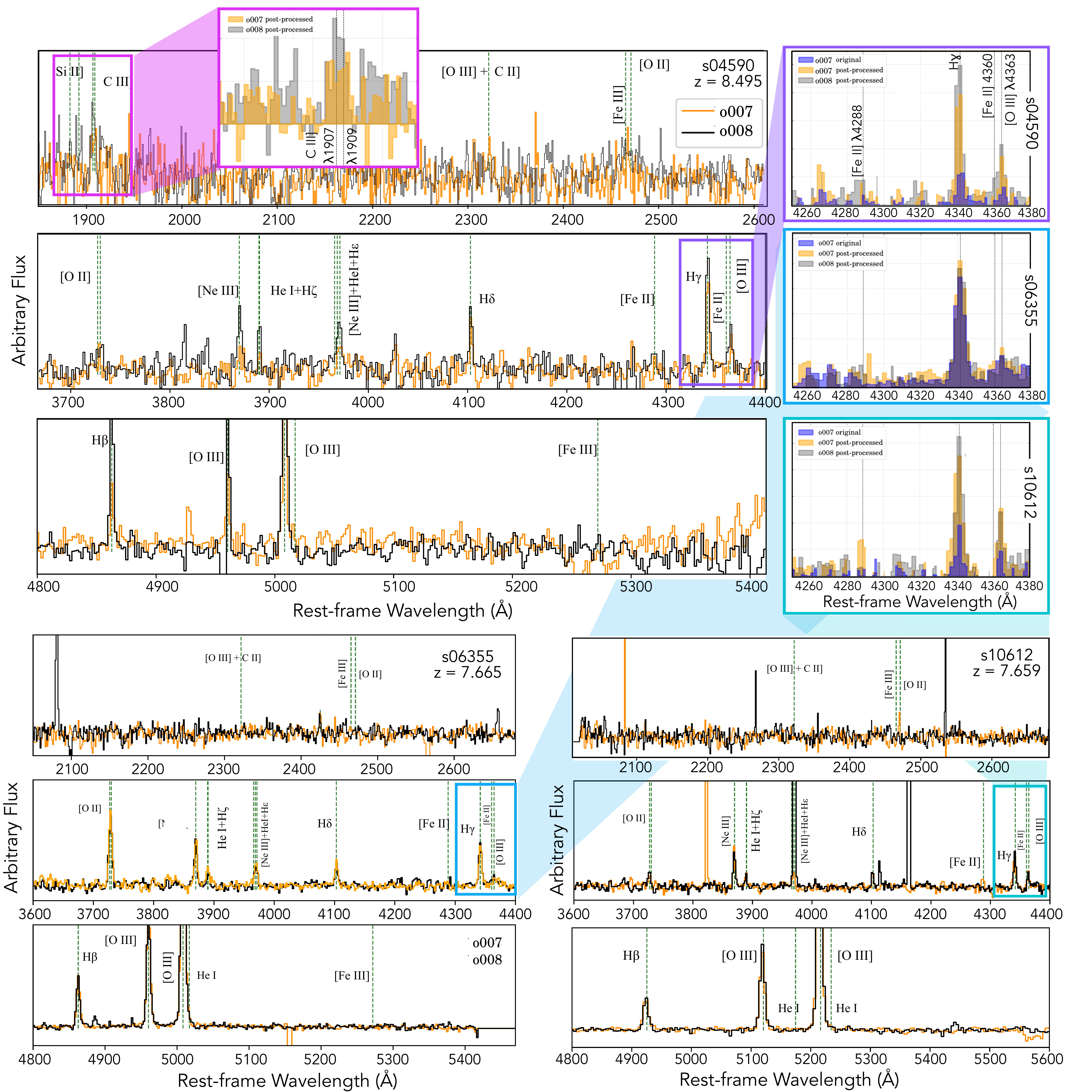}
    \caption{The JWST/NIRSpec post-processed rest-frame spectra of s04590, s06355, and s10612. 
    Wavelength ranges highlighting the important nebular emission lines 
    detections are shown. 
    The orange and black spectra correspond to the o007 and o008 visits, respectively, for each galaxy. 
    We highlight the detection of essential emission lines such as 
   \ion{C}{3}] \W\W1906,1906, [\ion{Ne}{3}] \W3869, 
    H$\gamma$ \W4101, H$\delta$ \W4340, [\ion{O}{3}] \W4363,  
    H$\beta$ \W4861, and [\ion{O}{3}] \W5007, and the tentative detection of [\ion{Fe}{3}] \W2465 in s04590.
    The inset spectra to the right highlight the differences between the original JWST Science Calibration Pipeline spectrum and the post-processed spectrum
    \citep[][]{trump22, arrabalharo22} for the H$\gamma$ and [\ion{O}{3}] \W4363 
    spectral features. 
    The orange and black spectra correspond to the two different post-processed spectra for the same galaxy, labeled as o007 and o008, respectively, and the original
    o007 reduction is shown in blue.
    For all three galaxies, the post-processed spectra show higher
    S/N in the emission lines.}
\label{fig:spectra}
\end{center}
\end{figure*}


\begin{deluxetable*}{l ccccc} 
\tablewidth{0pt}
\setlength{\tabcolsep}{4pt}
\tablecaption{Observed and corrected line intensities with respect to H$\beta$ = 100 for s04590, s0355, and s10612.}
\tablehead{
    &             \multicolumn{5}{c}{s04590}                                      \\
$\lambda_{0}$ &  Ion      & \mc{2}{o007$^{\star}$}  & \mc{2}{o008}   \\
 & &  $F$($\lambda$) & $I$($\lambda$)  & $F$($\lambda$) & $I$($\lambda$) \\
}      
\startdata
1909$^{\dagger}$        & [\ion{C}{3}]    &  53.1$\pm$22.2      &  212.56$\pm$95        &  21.6$\pm$7.8     &  86.4$\pm$34.0 \\
2465        & [\ion{Fe}{3}]   &  \nodata      &  \nodata       &   6.6$\pm$2.7        &  18.1$\pm$7.8    \\
3727        & [\ion{O}{2}]    &   21.9 $\pm$ 0.6    &  30.9$ \pm$ 1.51      &   15.9 $\pm$ 0.5     &  22.5 $\pm$ 1.1   \\
3869        & [\ion{Ne}{3}]   &   14.6 $\pm$ 0.4    &  19.9$\pm$0.91      &   26.7$\pm$0.8     &  36.5$\pm$1.7   \\
4101        &  H$\delta$      &   46.1$\pm$1.3    &  58.9 $\pm $2.35      &   20.7$ \pm$ 0.6     & $ 26.4 \pm$ 1.1   \\
4341        &  H$\gamma$      &   99.9 $\pm $2.8    &  118.3 $\pm$ 4.05     &   42.0$ \pm $1.2     &  49.7 $\pm $1.7   \\
4360$^{\ddagger}$  & [\ion{Fe}{2}]  & \nodata   &  \nodata  &  3.37$\pm$0.10  & 3.97$\pm$0.13  \\
4363        & [\ion{O}{3}]    &   33.9$ \pm$ 1.0    &  39.8 $\pm$ 1.34      &   12.1 $\pm$ 1.7     &  14.2 $\pm$ 2.0   \\
4959        & [\ion{O}{3}]    &   106.5$\pm$3.0   &  103.6$\pm$2.95     &   \nodata          &  \nodata        \\
5007        & [\ion{O}{3}]    &   366.1$\pm$10.4  &  351.6$\pm$10.08    &   330.2$\pm$9.3    &  317.1$\pm$9.1  \\
$F$(H$\beta$) &   &  1.18$\pm$0.02  & &  3.37$\pm$0.06  \\
\hline
\\
    &   \multicolumn{5}{c}{s06355} \\   
\multicolumn{1}{c}{$\lambda_{0}$} & \multicolumn{1}{c}{Ion} & \multicolumn{2}{c}{o007} & \multicolumn{2}{c}{o008} \\
 & &  $F$($\lambda$) & $I$($\lambda$)  & $F$($\lambda$) & $I$($\lambda$) \\
 \hline
3727        & [\ion{O}{2}]   & 76.4 $\pm$ 2.2   &  111.1$\pm$ 5.3   & 80.0  $\pm$ 2.3   &  105.1 $\pm$5.3  \\
3869        & [\ion{Ne}{3}]  & 39.0 $\pm$ 1.1    &  54.6 $\pm$ 2.5  & 42.0  $\pm$ 1.2   &  53.7  $\pm$2.5  \\
4101        &  H$\delta$     & 20.3 $\pm$ 0.6   &  26.5 $\pm$ 1.1   & 21.7  $\pm$ 0.6   &  26.4  $\pm$1.1  \\
4341        &  H$\gamma$     & 40.2 $\pm$ 1.1   &  48.2 $\pm$ 1.6   & 43.6  $\pm$ 1.2   &  49.8  $\pm$1.7  \\
4363        & [\ion{O}{3}]   & 11.6 $\pm$ 0.3   &  13.8 $\pm$ 2.0   & 9.6   $\pm$ 1.7   &  10.9  $\pm$2.0   \\
4959        & [\ion{O}{3}]   & 245.2$\pm$ 6.9   &  238.0$\pm$ 6.8   & 257.0 $\pm$ 7.3   &  251.4 $\pm$7.2  \\
5007        & [\ion{O}{3}]   & 822.9 $\pm$ 23.3  &  787.7$\pm$ 22.6 & 834.3 $\pm$ 23.6  &  808.1 $\pm$23.2 \\
\\
$F$(H$\beta$) &   &  2.24$\pm$0.04  & &  2.27$\pm$0.05  \\
\hline
\\
    &   \multicolumn{5}{c}{s10612} \\ 
\multicolumn{1}{c}{$\lambda_{0}$} & \multicolumn{1}{c}{Ion} & \multicolumn{2}{c}{o007} & \multicolumn{2}{c}{o008} \\
 & &  $F$($\lambda$) & $I$($\lambda$)  & $F$($\lambda$) & $I$($\lambda$) \\
\hline
3727        & [\ion{O}{2}]     &  14.8 $\pm$ 0.4      &  20.83 $\pm$1.0    &  32.4  $\pm$ 0.9  &  41.1  $\pm$2.0   \\
3869        & [\ion{Ne}{3}]   &  45.8 $\pm$ 1.3      &  62.38 $\pm$2.8    &  52.9  $\pm$ 1.5  &  65.7  $\pm$3.0   \\
4101        &  H$\delta$      &  20.7 $\pm$ 0.6      &  26.4  $\pm$1.0    &  22.3  $\pm$ 0.6  &  26.4  $\pm$1.0   \\
4341        &  H$\gamma$  &  49.8 $\pm$ 1.4      &  58.8  $\pm$2.0    &  48.6  $\pm$ 1.4  &  54.6  $\pm$1.9   \\
4363        & [\ion{O}{3}]     &  22.4 $\pm$ 0.6      &  26.21 $\pm$2.0    &  18.1  $\pm$ 1.8  &  20.2  $\pm$2.0   \\
4959        & [\ion{O}{3}]    &  209.2 $\pm$ 5.9     &  203.61 $\pm$ 5.8  &  248.6 $\pm$ 7.0  &  243.9 $\pm$6.9  \\
5007        & [\ion{O}{3}]    &  666.5 $\pm$ 18.9  &  640.44 $\pm$ 18.4 &  730.0 $\pm$ 20.6 &  709.9 $\pm$20.3 \\
\\
$F$(H$\beta$) &   &  1.37$\pm$0.03  & &  1.28$\pm$0.03  \\
\enddata
\tablecomments{$^{\star}$The o007 spectrum of s04590 is affected by shutter closures \citep[see][]{curti22a}. $^\dagger$\ion{C}{3}] \W1907+[\ion{C}{3}] \W1909. $^{\ddagger}$Excess blue profile in [\ion{O}{3}] \W4363 associated with the possible detection of [\ion{Fe}{2}] \W4360 in s04590 (see Fig.~\ref{fig:spectra}). The observed H$\beta$ fluxes (in units of 10$^{-16}$ erg s$^{-1}$ cm$^{-2}$ \AA$^{-1}$).}
\label{tab:intensities}
\end{deluxetable*}

\section{Electron temperature and Metallicity determinations} 
\label{sec:4}
The gas-phase metallicity of a galaxy is a key piece in the puzzle to 
understanding its evolution. 
Accurate abundance determinations require knowing the physical gas conditions.
While nebular abundances are insensitive to the typical range of observed electron
densities ($10^1<n_e (\mbox{cm}^{-3})<10^3$), nebular emission lines are exponentially
dependent on the electron temperature.
For high ionization galaxies, such as the $z>7$ NIRSpec galaxies investigated here ( see Sec. \ref{ion_parameter}),
a measure of the high-ionization temperature via the [\ion{O}{3}] \W4363/\W5007
ratio is needed.
Here we investigate the intrinsically-faint, temperature-sensitive [\ion{O}{3}] 
$\lambda$4363 auroral line in s04950, s06355, and s01612.
The detection of this line suggests that detailed, cosmic abundance determinations 
will be possible with JWST for more distant galaxies than ever before.
Below, we use the \texttt{PyNeb} package \citep[version 1.1.14;][]{luridiana15} 
in \texttt{Python} to calculate the physical conditions and chemical abundances in 
s04950, s06355, and s10612. 
We follow the same procedure and atomic data described in \citet{berg21a} and \citet{arellanocordova22}. For Ne and C, 
we have used the atomic data reported in \citet{arellano-cordova2020b} and \citep{berg19a}, respectively.

\subsection{The $T_e$-sensitive [\ion{O}{3}] \W4363 line} 
\label{sec:artifact}
A number of low- and high-ionization emission lines are evident in the JWST/NIRSpec 
spectra of s04950, s06355, and s01612.
Given the high redshift of these galaxies, the detection of weak, high-ionization 
emission lines such as the [\ion{O}{3}] \W4363 that allow the computation of the 
electron temperature of the nebular gas (see Figure~\ref{fig:spectra}) is spectacular.
We used the [\ion{O}{3}] \W4363/\W5007 line ratio from the post-processed spectra 
presented in \citet{trump22} to calculate the electron temperature in the $z>7$ galaxies. 
For the visit o008 of  s04590, we calculate a very high electron 
temperature that is near the H-cooling limit of  2.41$\times10^4$ K, while that for the o007 visit we assume $T_{\rm e}$ = 16000 K (which is a representative value of local-star-forming galaxies) due to the nonphysical value of $T_{\rm e}$([\ion{O}{3}]) determined using the [\ion{O}{3}] \W4363/\W5007 line ratio ~\citep[see also][]{schaerer22}. 
For s06335 and s10612, we calculate $T_{\rm e}$([\ion{O}{3}]) values for both the 
o007 and o008 spectra, finding a $\Delta T_e = 1.5\times10^3$ K for s06355 and
$\Delta T_e = 4.8\times10^3$ K for s10612.
The calculated electron temperatures are reported in Table~\ref{tab:metal}.

To investigate the large differences in the calculated electron temperatures,
we compare the H$\gamma+$[\ion{O}{3}] \W4363 spectral range in the o007 and o008 visits of each galaxy in the inset panels of Figure~\ref{fig:spectra}. 
The narrow-aperture-extraction 1D spectrum used in this analysis 
\citep[][gray and orange spectra]{trump22, arrabalharo22}  has higher signal-to-noise
than the o007 original JWST Science Calibration Pipeline data (blue spectra). 
All three spectra display a clear detection of H$\gamma$ and [\ion{O}{3}] $\lambda$4363 for the narrow-aperture-extraction post-processed 1D spectra. 
Focusing on the post-processed spectra, 
there are noticeable differences between the o007 and o008 spectra of 
s04590.
While the profile shapes are consistent between the o007 and o008 visits 
of s06355 and s10612, the [\ion{O}{3}] \W4363 profile shapes of s04590 differ.
Specifically, the o008 visit of s04590 shows excess flux blueward of the 
[\ion{O}{3}] \W4363 line center.

The source of the excess blue [\ion{O}{3}] \W4363 flux in the s04590 o008 visit 
is uncertain.
[\ion{Fe}{2}] \W4360 emission may be strong enough
to contaminate [\ion{O}{3}] \W4360 at moderate metallicities 
\citep[e.g.,][]{curti+17,berg20,arellano-cordova20,Rogers21,rogers22}.
However, these authors have only observed significant [\ion{Fe}{2}]\W4360 in high 
metallicity $z=0$ environments (12$+$log(O/H) $\sim 8.4$).
How this applies to high-$z$ environments in still unclear.
In the case of s04590, the center of the excess blue flux aligns extremely well with the vacuum wavelength of the [\ion{Fe}{2}] \W4360 line.
Fitting for the potential [\ion{Fe}{2}] \W4360 line simultaneously with the 
[\ion{O}{3}] \W4363 line results in a 3-$\sigma$ [\ion{Fe}{2}] detection that is 
22\% the strength of the blended line flux.
If not corrected for, [\ion{Fe}{2}] contamination could lead to an overestimate of the $T_e$([\ion{O}{3}]).

Unfortunately, other evidence for significant Fe emission is weak or uncertain.
In the case of detectable [\ion{Fe}{2}] \W4360 emission, [\ion{Fe}{2}] \W4288 emission, 
which arises from the same orbital level as [\ion{Fe}{2}]\W4360 
\citep[see details in][]{mendez-delgado21}, is expected at higher flux.
As shown in the Figure \ref{fig:spectra} inset windows, weak [\ion{Fe}{2}] \W4288
maybe present, but only at the 1-$\sigma$ level in the current s04590 visits.
On the other hand, the detection of the excess blue flux in [\ion{O}{3}]~\W4363 is only visible in the o008 spectrum, which might indicate the presence of a possible artifact affecting the flux and profile of [\ion{O}{3}]~\W4363.
This emphasizes that the current JWST observations present significant instrumental artifacts that must be accounted to produce consistent and reliable results. For our calculations, we have used the deblended flux measurement of [\ion{O}{3}] \W4363 to calculate $T_e$([\ion{O}{3}]) for the o008 visit of s04590.

\subsection{Total and Relative Abundances}
\subsubsection{12+log(O/H)}
\label{sec:O/H}
The most common method of directly determining O/H abundances involves using $T_e$[\ion{O}{3}] measurements and reddening-corrected line fluxes to calculate their respective ionic abundances, summing these ionic abundances, and correcting for any significant unseen ionization states.
In typical \ion{H}{2} regions, O$^+$ and O$^{+2}$ are the dominant ions and can be 
determined from the [\ion{O}{2}] \W3727 and [\ion{O}{3}] \W\W4959,5007 emission lines. 
In fact, even in extreme emission line galaxies, which show strong detections of 
very-high-ionization emission lines (e.g., \ion{C}{4} \W\W1548,1550, \ion{He}{2} 
\W\W1640,4686, [\ion{Fe}{5}] \W4227, [\ion{Ar}{4}] \W\W4711,4740), the O$^0$ and O$^{+3}$ 
ions each compose $\lesssim4$\% of the total O ions, and thus can be ignored. 
Therefore, we calculate O$^{+2}$/H$^+$ using the $T_e$[\ion{O}{3}] measurements and
O$^+$/H$^+$, using the \citet{garnett92} $T_e-T_e$ relations to estimate $T_e$[\ion{O}{2}],
and add these ionic abundances together to determine the total O/H abundance.
The resulting 12+log(O/H) abundances are reported in Table~\ref{tab:metal}.
In general, the metallicities for the $z>7$ galaxies are metal-poor 
(12+log(O/H) $< 8.2$ or Z$_{neb.} \lesssim 0.3 $Z$_\odot$).
To explore the impact of the highly uncertain reddening correction, we also calculated the metallicity using uncorrected line fluxes ($E(B-V)$ = 0), finding the O/H abundances decrease by 0.12 dex on average, 
which is within the uncertainties estimated in this study \citep[see also][]{curti22a}.

\subsubsection{Ionization Parameter}\label{ion_parameter}
Another key nebular property is the ionization parameter, log$U$,
which provides a measure of the strength of the ionizing radiation field.
Here we use a common proxy for ionization parameter, $O_{32}$,
defined by the [\ion{O}{3}] \W5007/\W3727 ratio. 
The resulting $O_{32}$ values for the present sample span a large range of 
high to very high ratios with $O_{32} \approx 7$ for s06335,
$O_{32} \approx 17$ for s04590, and 
$O_{32} \approx 30$ for s10612. 
Such high values of $O_{32}$ are often associated with extreme emission line galaxies
and/or Lyman continuum escape in the nearby universe  
\citep[e.g.,][]{izotov16, senchyna17, berg21a}. 

\subsubsection{Relative C/O and Ne/O Abundances}
The high-redshift of the s04590 spectrum allows a measure of the \ion{C}{3}] \W\W1907,1909
flux and, subsequently, a measure of the relative C$^{+2}$ ionic abundance.
Owing to the similar ionization and excitation energies of the C$^{+2}$ and O$^{+2}$
ions, relative C/O abundances are typically determined from the C$^{+2}$/O$^{+2}$ ratio. 
For high-ionization nebulae, such as the $z>7$ galaxies investigated here, 
we must also consider carbon contributions from the C$^{+3}$ species to avoid 
underestimating the true C/O abundance. 
To account for the potential contribution from C$^{+3}$, we use the
photoionization-model-derived C ionization correction factor (ICF) of \citet{berg19b}, which are determined using the ionization parameter.
In turn, we determine the ionization parameter based on the log([\ion{O}{3}] \W 5007/[\ion{O}{2}] \W3727) ratio and the relationship provided by \citet{berg19b}.

For Ne, we have used the high-ionization [\ion{Ne}{3}] \W3869 line to calculate the ionic 
abundances of Ne$^{+2}$/O$^{+2}$. 
To account for unseen ions of Ne, we use the ICF provided by \citet{amayo21}.
Similar to the method for C, this ICF is based on photoionization models and is determined as a function of the ionization parameter  ([\ion{O}{3}]~$\lambda$5007/[\ion{O}{2}]~$\lambda$3727).
\citet{amayo21} provide relationships for the NE ICF as a function of O$^{2+}$/(O$^{+}+$O$^{2+}$) ratio to characterize the average ionization within the nebula. 
We list the resulting Ne/O abundances for each galaxy and the C/O abundance for J04590 in Table~\ref{tab:metal}.

\begin{deluxetable*}{l cccc |  cccc | cccc} 
\tablewidth{0pt}
\setlength{\tabcolsep}{4pt}
\tablecaption{Derived galaxy properties, metallicities and ionization parameter}
\tablehead{ 
{}                & \mc{2}{s04590}                    &&& \mc{2}{s06335}                  &&& \mc{2}{s10612}                                    \\
[0ex]
Property     & {\it o007\sn}   &   o008  &  \D &     & o007 &   o008   & \D   &&    o007  &   o008 & \D }     
\startdata
$z$                   & 8.495       & 8.495        &      && 7.665        & 7.665       &      && 7.659        & 7.659        &      \\
$E(B-V)$          & {\it 0.32}  & 0.32$\pm$0.04         &      && 0.34$\pm$0.04          & 0.26$\pm$0.04         & 0.08 && 0.31$\pm$0.04          & 0.22$\pm$0.04          & 0.09 \\
$T_e$[\ion{O}{3}] & \ND   & 2.41$\pm$0.25  & \ND && 1.44$\pm$0.10  & 1.29$\pm$0.10 & 0.15 && 2.28$\pm$0.90  & 1.80$\pm$0.10  & 0.48 \\   
O$_{32}$          & {\it 11.4}  & 14.1         & 2.7  && 7.1          & 7.7         & 0.6  && 30.1         & 17.3         & 12.8 \\
12$+$log(O/H)     & {\it 7.54}  & 7.12$\pm$0.12  & 0.42 && 8.03$\pm$0.12  & 8.17$\pm$0.12 & 0.14 && 7.45$\pm$0.07  & 7.73$\pm$0.08  & 0.28 \\
log(C/O)          & {\it $-$0.23} & -0.83$\pm$0.38 & 0.60 && \ND          & \ND         & \ND  && \ND          & \ND          & \ND  \\
log(Ne/O)         & {\it $-$0.76}    &  -0.52$\pm$0.15 & 0.24 && -0.64$\pm$0.13 &-0.64$\pm$0.17 & 0.00 && -0.58$\pm$0.06 & -0.58$\pm$0.09 & 0.00  
\enddata
\tablecomments{
For each galaxy we list the properties from Column 1 for both the o007 and o008 visits, 
followed by the difference between each pair of measurements.
Row 1 provides the redshifts measured in this work.
Rows 2--4 list our calculated nebular properties for reddening, $E(B-V)$,
electron temperature, $T_e$  (10$^4$ K), and ionization, as represented by the 
[\ion{O}{3}] \W5007/[\ion{O}{2}]\W3727 line ratio.
The reddening values were calculated using the H$\delta$/H$\beta$ ratio and the \citet{cardelli89} reddening law.
Electron temperatures were calculated from the [\ion{O}{3}] \W4363/\W5007 line ratio and
are given in units of $10^4$ K.
Rows 5--7 list the individual values and comparative differences
for the derived abundances: 12+log(O/H), log(C/O), and log(Ne/O). 
$^\star$ Affected by shutter closures \citep[see][]{curti22a}. Assuming $T_e$ = 16000 K to calculate the chemical abundances.}
\label{tab:metal}
\end{deluxetable*}

\section{Discussion}\label{sec:discussion}
\subsection{The Mass-metallicity relation }
The relationship between stellar mass and gas-phase metallicity (the MZR) 
provides a critical probe of galaxy evolution. 
The MZR is shaped by the cumulative galaxy evolutionary processes, 
where star formation builds up metals through nucleosynthesis, 
while enriched outflows remove metals and pristine gas inflows dilute the metals in the ISM 
\citep[e.g.,][]{tremonti04, dalcanton07, peeples11, zahid14}.
Because accurate MZR trends require direct abundance measurements, 
the evolution of the MZR has only been investigated at moderate redshifts
\citep[e.g.,][]{sanders22} and typically only probes the most massive systems. 
The advent of JWST opens a window onto the distant universe
that pushes the MZR to higher redshifts and lower masses
than ever before. 

\begin{figure}
\begin{center}
    \includegraphics[width=0.45\textwidth, trim=30 0 30 0,  clip=yes]{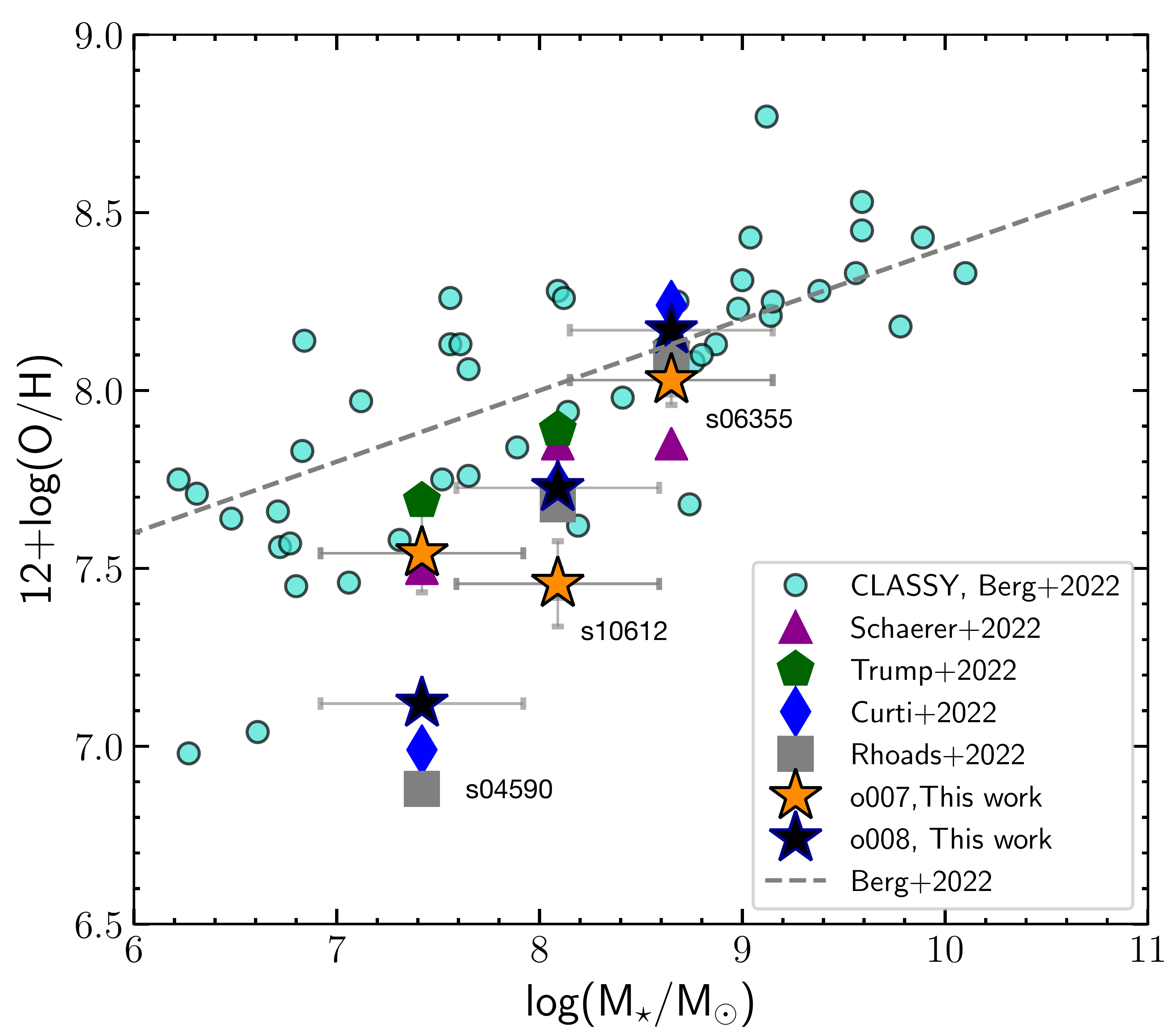}
    
    \caption{The stellar mass versus gas-phase metallicity (12+log(O/H)) relationship for star-forming galaxies. 
    The $z>7$ galaxies are plotted as stars with stellar masses derived by \citet{carnall22}. 
    Additional recent metallicity determinations are labeled from \citet[][triangles]{schaerer22}, 
    \citet[][pentagons]{trump22}, 
    \citet[][diamonds]{curti22a}, and
    \citet[][squares]{rhoads22}. 
    For the purpose of comparing to local galaxy populations, 
    we plot the UV-bright CLASSY sample \citep[circles+dashed line;][]{berg22}}
\label{fig:spec-mass-metal}
\end{center}
\end{figure}

In Figure~\ref{fig:spec-mass-metal}, we use our O/H abundance determinations with the stellar masses reported by \citet{carnall22} to plot the MZR for 
the three $z>7$ galaxies of JWST/NIRSpec galaxies.
Following \citet{trump22}, we adopted large (0.5 dex) error bars 
to account for the systematic offset of the \citet{carnall22} masses from 
the \citet{schaerer22} and \citet{curti22a} masses.
Figure~\ref{fig:spec-mass-metal} compares other recent analyses of these galaxies from \citet{schaerer22}, \citet{trump22}, \citet{curti22a}, and \citet{rhoads22}. 
Note that the results of \citet{schaerer22} and \citet{rhoads22} are 
obtained using the original JWST Science Calibration Pipeline spectra, 
while \citet{trump22} and \citet{curti22a} used their own post-processing
reduction.
All four studies determined abundances from the combined o007 and o008 spectra for each galaxy\footnote{Note that \citet{curti22a} took care to remove a nod of the \textit{o007} visit for s04590 that was affected by a
failed shutter.},
Additionally, each of these studies used different methods to calculate metallicity \citep[see e.g.,][]{trump22}.
In general, we find that our O/H measurements for s06335 and s10612 are
consistent with the literature determinations, with an
overall dispersion of $\sigma\sim0.5$~dex.
On the other hand, the abundances for s04590 are separated into two distinct
clumps in Figure~\ref{fig:spec-mass-metal}, which could be due to the odd line profile between the two different visits, indicating the sensitivity of the $T_e$ determination to uncertainties in the [\ion{O}{3}] \W4363 flux \citep{arellanocordova16}.
However, a definitive interpretation is complicated due to the differences in reduction and calculation methods.

To investigate the evolution of the MZR with redshift, 
Figure~\ref{fig:spec-mass-metal} compares the $z>7$ galaxies to the $z\sim0$ MZR trends from the COS Legacy Archive Spectroscopic SurveY 
\citep[CLASSY;][]{berg22}.
While for s06355 and s10612, the values of metallicity are nearly consistent with previous determinations within the uncertainties, the results for s04590 show significant differences, which are divided into two different ranges of metallicity (see Figure~\ref{fig:spec-mass-metal}).
In particular, if the higher abundance of s04590 is adopted, the slope of the $z>7$ is
consistent with the $z\sim0$ trend, suggesting little redshift evolution. This would suggest that the physical properties that set the MZR at $z\sim0$
(outflows of metal-enriched gas and inflows of pristine gas) are already in place
by $z\sim7$, and their relative strengths do not significantly evolve with time. 
Alternatively, if we adopt the lower s04590 abundance, the $z>7$ MZR has a much steeper slope.
The slope of the MZR is largely shaped by the level of feedback in galaxies, where
a steeper slope would imply either
(1) an increased importance of inflows of pristine gas at lower-masses / metallicities and/or
(2) weaker outflows of enriched gas, allowing the O/H abundances to more quickly build up with stellar mass than seen at $z\sim0$.
Although the analysis based on the result of s04590 might suggest a change in the slope of the MZR at $z > 7$, its position in the MZR is still consistent with the observed values at $z\sim$0 due to the dispersion observed in $z\sim0$ galaxies \citep[][$\sigma$ = 0.29 dex]{berg22}.
Then, the interpretation of the MZR at $z > 7$ strongly depends on the data quality/reductions and the sample size. Therefore, large samples of $z>7$ galaxies  with high S/N and robust absolute flux calibration are needed to further constrain the evolution of the MZR.

\subsection{Ionization Parameter Evolution}
Table~\ref{tab:metal} lists the O$_{32}$ values measured from the o007 and o008 visits for each galaxy. 
The ionization of the $z>7$ sample is high on average: $\langle$O$_{32}\rangle = 15$,
which is consistent with studies of nearby extreme emission line galaxies that are typical of LyC escape \citep{izotov12, izotov16, senchyna19}, but much higher than the general $z\sim0$ galaxy population.
In comparison to the recent literature, 
we find that the O$_{32}$ value of o007 for s10612 is consistent with the result provided 
by \citet{curti22a} (O$_{32} = 27.3$), but the o008 visit has a much lower value
of O$_{32} = 17.3$.
Additionally, our O$_{32}$ values are much higher in general than those  
reported by \citet{schaerer22}, especially for s04590 and s10612. 
The differences in reported O$_{32}$ values are likely due to the issues with the 
current pipeline-produced 1D spectra (v1.6.1) and the inferred dust correction. 
This suggests that the O$_{32}$ value strongly depends on the adopted reduction routine (e.g., the narrow-extraction aperture) and may dramatically change as the NIRSpec reference files and reduction routines are refined.

\subsection{The C/O--O/H and Ne/O--O/H Trends}
\begin{figure*}
\begin{center}
    \includegraphics[width=1.0\textwidth, trim=0mm 0mm 0mm 0mm, clip=yes]{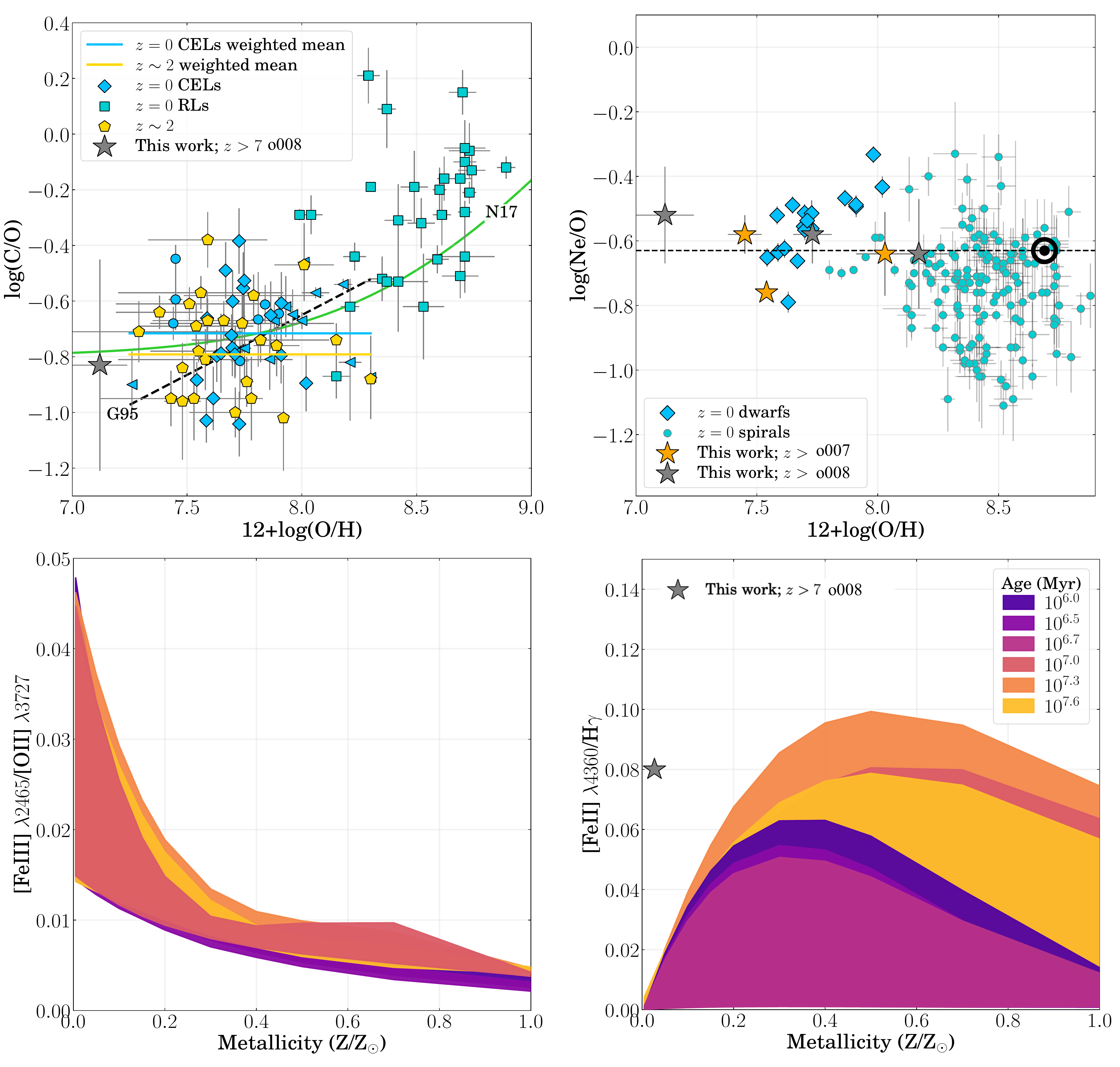}
    \caption{
    \textit{Top left:} The C/O versus O/H relationship taken from \citet{berg19a} for 
    $z\sim0$ and $z\sim2$ galaxies. The dashed and solid lines represent the relationships of \citet{garnett95} and \citet{nicholls17}. 
    The results for s04590 are shown as orange star for the o008 visit.
    \textit{Top right:} The Ne/O versus O/H relationship for $z\sim0$ 
    low-metallicity dwarf galaxies and \ion{H}{2} regions of spiral galaxies. 
    The Ne/O ratios for the $z>7$ NIRspec galaxies are displayed as gray and cyan symbols. 
    The black dashed line indicates the solar Ne/O value \citep[$-0.63\pm0.05$;][]{asplund21}. 
    \textit{Bottom row:} Photoionization models of the [\ion{Fe}{3}]\W2465 / [\ion{O}{2}] \W3727 
    flux (left) and [\ion{Fe}{2}] \W4360/H$\gamma$ flux (right) versus metallicity. The value derived for the \ion{Fe}{3}]\W2465 / [\ion{O}{2}] \W3727 ratio is outside the range of the models (bottom left panel).
    }
\label{fig:C-Ne}
\end{center}
\end{figure*}

We present the C/O versus O/H relationship for our $z>7$ sample in the top left panel of 
Figure~\ref{fig:C-Ne}. 
Unfortunately, only s04590 has a high enough redshift to measure the far-UV \ion{C}{3}] \W\W1907,1909 lines, and thus we only have C/O results for this galaxy. 
In comparison, we include the $z\sim0$ and $z\sim2$ trends from \citet{berg19a}.
For 12+log(O/H)$\lesssim8.0$, these trends derive C/O using the rest-frame far-UV 
\ion{O}{3}] \W1666 and \ion{C}{3}] \W\W1907,1909 collisionally-excited lines (CELs), 
while the C/O abundance at higher metallicities is derived using the rest-frame 
optical \ion{C}{2} \W4267 recombination line (RL).
This first-look and analysis of the log(C/O) abundance at $z>7$ (from the o008 visit) shows a value of log(C/O) lower than the average observed for $z\sim0$ and $z\sim 2$ galaxies. However, the log (C/O) values for $z \sim$0$-$2 galaxies show a large dispersion at low-metallicity, $\sigma \sim$ 0.17 dex \citep{berg19a}. Therefore, the log(C/O) value derived for s4590 remains consistent with the observed values in local star-forming galaxies.
Interestingly, from modeling the chemical evolution of C/O of metal-poor galaxies,
\citet{berg19a} found that the C/O ratio is very sensitive to both the detailed star 
formation history and supernova feedback. 
Along these lines, the $z>7$ C/O abundance plotted in Figure~\ref{fig:C-Ne} points to either lower star formation efficiencies,
a longer burst duration, a larger effective oxygen yield (little ejection of oxygen in outflows), or some combination of these. However, note that the C/O ratio for s04590 (see Table~\ref{tab:metal}) is highly uncertain, and robust measurements are required to determine what drives the chemical abundances at $z$ > 7. 
In Table~\ref{tab:metal}, we also provide the C/O ratio derived from the o007 spectrum. Such a value is significantly higher than the o008 visit and inconsistent with the $z\sim0$ and $z\sim2$ trends.
A large C/O value would be our first hint that the C/O--O/H relationship evolves significantly over cosmic time, where pristine inflows were larger and/or effective oxygen yields were lower at $z>7$. However, the detection of \ion{C}{3}] \W\W1907,1909 lines in the o007 visit are uncertain, and the C/O ratio depends of our assumption of $T_e = 16000 $ K.

In the right top panel of Figure~\ref{fig:C-Ne} we show the relation between 
log(Ne/O) and metallicity for our $z>7$ sample.
For a local comparison, we also plot a sample of dwarf galaxies \citep{berg19a} and \ion{H}{2} regions from spiral galaxies \citep{berg20}. 
In principle, Ne is an $\alpha$-element and is thus expected to be constant with O/H  
because massive stars produce both elements on the same time scales \citep[see e.g.,][]{arellano-cordova2020b, Rogers21}. 
For our $z>7$ sample, we find consistent Ne/O abundances with the $z\sim0$ 
low-metallicity dwarf-galaxies.
Despite the large observational uncertainties associated with this study (e.g., absolute flux calibration and possible artifacts in relevant emission lines),
all of the individual Ne/O measurements are consistent within 0.2 dex of a solar 
abundance \citep{asplund21}. 

In addition, to investigate the impact of the reddening in determining the Ne/O ratio, we derived the Ne/O abundance ratio using the intensities of [\ion{O}{2}], [\ion{O}{3}] and [\ion{Ne}{3}] and $T{_e}$[\ion{O}{3}] reported in \citet{curti22a}. The intensities were corrected by reddening using the $A_{v}$ values also reported in \citet{curti22a}. We calculated values of the Ne/O ratios of  $-0.71\pm0.25$, $-0.67\pm0.12$ and $-0.58\pm0.17$ for s04590, s06355, and s10612. Such results are consistent with respect to those derived in this work (see Table~\ref{tab:metal}). However, for the o008 visit of s04590, we reported a value 0.19 dex higher than that obtained using the measurements of \citet{curti22a} but consistent within the errors. 
Despite the different approaches used to derive reddening and T$_{e}$[\ion{O}{3}], both results are in agreement with the Ne/O abundance ratios estimated in local star-forming regions at low metallicity (12+log(O/H) < 8.0), and with respect to the Solar abundance as reported in local star-forming regions \citep[see e.g.,][]{Dors13}. 

Therefore, we find that Ne/O abundances can be robustly determined from JWST observations of high-redshift galaxies, and {at least in this analysis} there is no evidence 
for cosmic evolution of the Ne/O--O/H trend. 
We stress that this first look at the C and Ne lines and their respective ionic abundances and the Ne/O and C/O ratios provide a first reference of the expected values at $z>7$ compared to local galaxies, and the challenges in deriving robust measurements of other metals in addition to O/H.

\subsection{Fe/O Abundance at High Redshift}
While collisionally-excited emission lines are commonly observed for one or two 
species of Fe in $z=0$ \ion{H}{2} regions, Fe abundance determinations are often 
avoided due to the importance of dust depletion, accurate ICFs, and fluorescence \citep{rodriguez99, rodriguez02}. 
However, several recent studies have revived the interest in Fe abundances by suggesting 
that enhanced $\alpha$/Fe abundance ratios are responsible for the extremely hard 
radiation fields inferred from the stellar continua and emission-line ratios in 
chemically young, $z\sim2$ galaxies 
\citep[e.g.,][]{steidel18, shapley19, topping20}. 
Given the importance of $\alpha$/Fe (e.g., O/Fe) abundances to interpret the 
ionizing continua of early galaxies, we were motivated to investigate the Fe/O abundance in the $z>7$ galaxies.  

Perhaps unsurprisingly for chemically-young $z>7$ galaxies, we do not detect any significant Fe emission lines in the s06355 and s1016 spectra. The s04590 spectrum, on the other hand, shows tentative evidence of Fe emission
via the [\ion{Fe}{2}] \W\W4287,4360 and [\ion{Fe}{3}] \W\W2465,5271 lines. While robust Fe/O abundances cannot be produced from these lines alone \citep[see discussion in][]{berg21a}, we can gain some insight by examining the [\ion{Fe}{3}]/[\ion{O}{2}] line ratio,
whose ions have similar ionization potentials.
In the bottom panels of Figure~\ref{fig:C-Ne} we plot the photoionization grid
of \citet[][]{berg19}
of [\ion{Fe}{3}] \W2465/[\ion{O}{2}] \W3727 versus metallicity (left)
and [\ion{Fe}{2}] \W4360/H$\gamma$ (right).
BPASS singe-burst models with ages of [$10^{6.0}$,$10^{6.5}$,$10^{6.7}$, $10^{7.0}$, $10^{7.3}$, $10^{7.6}$] years were used as the input ionizing spectrum, as indicated by the different-colored shaded regions.
The vertical shading represents the maximum range of values produced considering a range of ionization parameters from log$U=-1$ to $-4$.
We over-plot the line ratios measured for the o008 visit of s04590,
however, the model ratios fail to reach the tentative detections of the line ratios for both 
the [\ion{Fe}{3}]/[\ion{O}{2}] and [\ion{Fe}{2}]/H$\gamma$ ratios. In particular, for the [\ion{Fe}{3}]/[\ion{O}{2}], we obtain a high value of such a ratio, which could be due to the high uncertainty in the detection of [\ion{Fe}{3}] in the o008 visit.
However, this mismatch may indicate that Fe/O abundances are {\it higher} at $z>7$ 
than $z\sim0$ (in contrast to the enhanced $\alpha$-Fe theory), 
which could result if high redshift galaxies have a low effective yield of oxygen, as suggested by the enhanced C/O abundance of the o007 visit of s04590.
At the same time, the current Fe detection is very tentative and may be overestimated due to the low-S/N < 3, especially given the lack of a secure flux calibration. Therefore, the Fe/O will need to be revisited.

\section{Conclusions} 
\label{sec:conclusion}

We analyze the rest-frame near-UV and optical spectra of three $z>7$ 
galaxies observed with Early Release Observations from 
NIRSpec on JWST. 
Each of the three galaxies (source IDs s04590, s06335, and s10612) were
observed for two visits (o007 and o008), allowing us to assess the 
quality of the spectra by examining the consistency of the emission features.
However, the default spectra from the JWST Science Calibration Pipeline (v1.6.1)
are not yet fully calibrated and, thus, the exact reduction process changes the delivered spectrum.
Of particular importance is the lack of a robust relative flux calibration,
the effects of failed shutter openings, and the potential
wavelength-dependent light loss during 1D extraction apertures 
\citep[see a complete discussion in ][]{trump22}. 
To mitigate some of these effects, our analysis used the post-processed JWST/NIRSpec spectra from 
\citet{trump22} and \citet{arrabalharo22}, which includes a custom 
aperture extraction and flux calibration. 
The resulting spectra still have large uncertainties and so we take extra caution
to use nearby (when possible) emission lines ratios to calculate the physical 
conditions, chemical abundances, and ionization parameters for the $z>7$ galaxies. 
Further, the o007 spectrum of s04590 is likely affected by a shutter failure,
causing a large discrepancy in its derived properties relative to the o008 spectrum.

With these data reduction caveats in mind, our main results are summarized as follows: 

$\bullet$ The temperature-sensitive [\ion{O}{3}] \W4363 line is detected in the two 
different visits (o007 and o008) of each high-$z$ galaxy, 
allowing the computation of $T_{e}$[\ion{O}{3}] for 5/6 of the individual spectra. 
However, we confirm the nonphysical result of $T_{e}$[\ion{O}{3}] derived for 
the o007 visit of s04590 (the visit with the failed shutter) reported by \citet{curti22a}. Based on the broad, asymmetric emission-line profile of the [\ion{O}{3}] \W4363
feature in the o008 visit of s04590, we explore
a possible reason for this result, indicating a potential contamination
of the line by either the [\ion{Fe}{2}] \W4360 line or a spectral artifact.

$\bullet$ Using our $T_{e}$[\ion{O}{3}] values, we derived direct-method O/H
abundances for the $z>7$ galaxies.
Recent studies from \citet{schaerer22,trump22,curti22a,rhoads22} have also 
reported metallicity measurements, but all use spectra from combined visits and
use a variety of analysis methods.
We find our measurements for s06335 and s10612 to be consistent within 2$\sigma$
between o007 and o008 and of those recently published in the literature.
For s04590, however, we find our two measurements to be discrepant by 0.42 dex,
where the o007 measurement (the visit with the failed shutter) is consistent with the studies of \citet{trump22} and \citet{schaerer22} and
the o008 is well-aligned with the values found by \citet{curti22a} and \citet{rhoads22}.
We use our O/H results to investigate the mass-metallicity relationship (MZR),
finding that the s04590 measurement changes the $z>7$ trend and that its large 
uncertainty significantly impacts our interpretation of the MZR:
(1) If the o007 (higher) abundance is adopted for s04590 (the visit with the failed shutter), then the $z>7$ MZR is consistent with
the $z>0$ trend.
(2) If the o008 (lower) abundance is adopted, the slope of the $z>7$ MZR becomes much steeper
than the $z>0$ MZR, suggesting different physical processes drive the shape of the MZR at $z>7$ that allow the 
O/H abundance to build up more quickly. However, due to the significant variation of the metallicity in s04590 derived in both visits and other studies from the literature, a detailed physical interpretation of the MZR cannot be done until larger samples of higher quality data are properly calibrated.

$\bullet$ At the redshift of s04590 ($z = 8.495$), the NIRSpec spectrum covers the \ion{C}{3}] \W\W1907,1909 lines. We measured this feature and used it to estimate the relative C/O abundance for s04590;
the most distant C/O measurement to date. 
When compared to $z\sim0-2$ low-metallicity dwarf galaxies, the o008 measurement is consistent given the large dispersion of the logC/O values of $z\sim0-2$ galaxies, and therefore, there is no evidence of an evolution in the C/O versus O/H relationship.

$\bullet$ We measured the first Ne/O abundance for $z>7$ galaxies using the isolated
[\ion{Ne}{3}] \W3868 emission line, which is detected in all three galaxies.
Our Ne/O abundances are consistent with a constant trend with O/H, as
expected for $\alpha$-element abundance ratios.
Our results are also in agreement with $z\sim0$ low-metallicity galaxies,
indicating that there is no redshift evolution of the Ne/O abundance.

$\bullet$ We analyzed the tentative detection of [\ion{Fe}{2}] \W4360 and [\ion{Fe}{3}] \W2465 in s04590. However, these flux measurements should be taken with caution.   
We find an [\ion{Fe}{3}]/[\ion{O}{2}] 
ratio is large compared to $z\sim0$ photoionization models, probably due the high uncertain in the measurement of [\ion{Fe}{3}] \W2465.
On the other hand, our simultaneous fit of the potentially blended [\ion{Fe}{2}] \W4360 $+$
[\ion{O}{3}] \W4363 lines yields an [\ion{Fe}{2}] \W4360/H$\gamma$ ratio that is only reproduced by the photoionization models at higher metallicities.
This may indicate significant Fe enrichment relative to O. We postulate that massive outflows of O-enriched gas at very early cosmic times could decrease the O-abundance relative to Fe to produce these large flux ratios. However, the tentative detection of Fe lines in s04590 should be revised, and upcoming JWST data with robust flux calibration and high S/N observation will allow us to assess the detection of such lines at $z > 7 $ galaxies.

In summary, JWST/NIRSpec spectra of high-redshift galaxies is opening an important window
on to the evolution of the first galaxies, such as this first look at the abundance
patterns of $z>7$ galaxies.
Thus, significant advancements in this area are imminent, but a robust interpretation will require 
larger samples of high-redshift galaxy observations, higher S/N spectra, and further 
refinement of the spectra reduction.

We thank the referee for thoughtful feedback that improved this letter.
The Early Release Observations and associated materials were developed, executed, and compiled by the ERO production team:  Hannah Braun, Claire Blome, Matthew Brown, Margaret Carruthers, Dan Coe, Joseph DePasquale, Nestor Espinoza, Macarena Garcia Marin, Karl Gordon, Alaina Henry, Leah Hustak, Andi James, Ann Jenkins, Anton Koekemoer, Stephanie LaMassa, David Law, Alexandra Lockwood, Amaya Moro-Martin, Susan Mullally, Alyssa Pagan, Dani Player, Klaus Pontoppidan, Charles Proffitt, Christine Pulliam, Leah Ramsay, Swara Ravindranath, Neill Reid, Massimo Robberto, Elena Sabbi, Leonardo Ubeda. The EROs were also made possible by the foundational efforts and support from the JWST instruments, STScI planning and scheduling, and Data Management teams.\\
The CEERS team thanks Pierre Ferruit and the NIRSpec GTO team for providing NIRSpec IPS simulated data and for general good counsel, and the STScI NIRSpec instrument team for extensive assistance regarding the \textit{JWST} Pipeline and data simulations.


\bibliography{mybib}

\begin{thebibliography}{}
\expandafter\ifx\csname natexlab\endcsname\relax\def\natexlab#1{#1}\fi
\providecommand{\url}[1]{\href{#1}{#1}}
\providecommand{\dodoi}[1]{doi:~\href{http://doi.org/#1}{\nolinkurl{#1}}}
\providecommand{\doeprint}[1]{\href{http://ascl.net/#1}{\nolinkurl{http://ascl.net/#1}}}
\providecommand{\doarXiv}[1]{\href{https://arxiv.org/abs/#1}{\nolinkurl{https://arxiv.org/abs/#1}}}

\bibitem[{{Amayo} {et~al.}(2021){Amayo}, {Delgado-Inglada}, \&
  {Stasi{\'n}ska}}]{amayo21}
{Amayo}, A., {Delgado-Inglada}, G., \& {Stasi{\'n}ska}, G. 2021, \mnras, 505,
  2361, \dodoi{10.1093/mnras/stab1467}

\bibitem[{{Arellano-C{\'o}rdova} {et~al.}(2020){Arellano-C{\'o}rdova},
  {Esteban}, {Garc{\'\i}a-Rojas}, \&
  {M{\'e}ndez-Delgado}}]{arellano-cordova2020b}
{Arellano-C{\'o}rdova}, K.~Z., {Esteban}, C., {Garc{\'\i}a-Rojas}, J., \&
  {M{\'e}ndez-Delgado}, J.~E. 2020, \mnras, 496, 1051,
  \dodoi{10.1093/mnras/staa1523}

\bibitem[{{Arellano-C{\'o}rdova} \& {Rodr{\'i}guez}(2020)}]{arellano-cordova20}
{Arellano-C{\'o}rdova}, K.~Z., \& {Rodr{\'i}guez}, M. 2020, \mnras, 497, 672,
  \dodoi{10.1093/mnras/staa1759}

\bibitem[{{Arellano-C{\'o}rdova} {et~al.}(2016){Arellano-C{\'o}rdova},
  {Rodr{\'i}guez}, {Mayya}, \& {Rosa-Gonz{\'a}lez}}]{arellanocordova16}
{Arellano-C{\'o}rdova}, K.~Z., {Rodr{\'i}guez}, M., {Mayya}, Y.~D., \&
  {Rosa-Gonz{\'a}lez}, D. 2016, \mnras, 455, 2627,
  \dodoi{10.1093/mnras/stv2461}

\bibitem[{{Arellano-C{\'o}rdova} {et~al.}(2022){Arellano-C{\'o}rdova},
  {Mingozzi}, {Berg}, {James}, {Rogers}, {Aloisi}, {Amor{\'\i}n}, {Brinchmann},
  {Charlot}, {Chisholm}, {Heckman}, {Dub{\'o}n}, {Hayes}, {Hernandez}, {Jones},
  {Kumari}, {Leitherer}, {Martin}, {Nanayakkara}, {Pogge}, {Sanders},
  {Senchyna}, {Skillman}, {Stark}, {Wofford}, \& {Xu}}]{arellanocordova22}
{Arellano-C{\'o}rdova}, K.~Z., {Mingozzi}, M., {Berg}, D.~A., {et~al.} 2022,
  \apj, 935, 74, \dodoi{10.3847/1538-4357/ac7854}

\bibitem[{{Arrabal Haro} {et~al.}(in preparation)}]{arrabalharo22}
{Arrabal Haro}, P., {et~al.} in preparation, \apj

\bibitem[{{Asplund} {et~al.}(2021){Asplund}, {Amarsi}, \&
  {Grevesse}}]{asplund21}
{Asplund}, M., {Amarsi}, A.~M., \& {Grevesse}, N. 2021, \aap, 653, A141,
  \dodoi{10.1051/0004-6361/202140445}

\bibitem[{{Berg} {et~al.}(2019{\natexlab{a}}){Berg}, {Chisholm}, {Erb},
  {Pogge}, {Henry}, {et~al.}}]{berg19b}
{Berg}, D.~A., {Chisholm}, J., {Erb}, D.~K., {et~al.} 2019{\natexlab{a}},
  \apjl, 878, L3, \dodoi{10.3847/2041-8213/ab21dc}

\bibitem[{{Berg} {et~al.}(2021){Berg}, {Chisholm}, {Erb}, {Skillman}, {Pogge},
  \& {Olivier}}]{berg21a}
---. 2021, arXiv e-prints, arXiv:2105.12765.
\newblock \doarXiv{2105.12765}

\bibitem[{{Berg} {et~al.}(2019{\natexlab{b}}){Berg}, {Erb}, {Henry},
  {Skillman}, \& {McQuinn}}]{berg19}
{Berg}, D.~A., {Erb}, D.~K., {Henry}, R.~B.~C., {Skillman}, E.~D., \&
  {McQuinn}, K.~B.~W. 2019{\natexlab{b}}, \apj, 874, 93,
  \dodoi{10.3847/1538-4357/ab020a}

\bibitem[{{Berg} {et~al.}(2019{\natexlab{c}}){Berg}, {Erb}, {Henry},
  {Skillman}, \& {McQuinn}}]{berg19a}
---. 2019{\natexlab{c}}, \apj, 874, 93, \dodoi{10.3847/1538-4357/ab020a}

\bibitem[{{Berg} {et~al.}(2020){Berg}, {Pogge}, {Skillman}, {Croxall},
  {Moustakas}, {Rogers}, \& {Sun}}]{berg20}
{Berg}, D.~A., {Pogge}, R.~W., {Skillman}, E.~D., {et~al.} 2020, \apj, 893, 96,
  \dodoi{10.3847/1538-4357/ab7eab}

\bibitem[{{Berg} {et~al.}(2013){Berg}, {Skillman}, {Garnett}, {Croxall},
  {Marble}, {et~al.}}]{berg13}
{Berg}, D.~A., {Skillman}, E.~D., {Garnett}, D.~R., {et~al.} 2013, \apj, 775,
  128

\bibitem[{{Berg} {et~al.}(2022){Berg}, {James}, {King}, {Mcdonald}, {Chen},
  {Chisholm}, {Heckman}, {Martin}, {Stark}, {The Classy Team}, {:}, {Aloisi},
  {Amor{\'I}n}, {Arellano-C{\'O}rdova}, {Bayliss}, {Bordoloi}, {Brinchmann},
  {Charlot}, {Chevallard}, {Clark}, {Erb}, {Feltre}, {Hayes}, {Henry},
  {Hernandez}, {Jaskot}, {Jones}, {Kewley}, {Kumari}, {Leitherer}, {Llerena},
  {Maseda}, {Mingozzi}, {Nanayakkara}, {Ouchi}, {Plat}, {Pogge},
  {Ravindranath}, {Rigby}, {Sanders}, {Scarlata}, {Senchyna}, {Skillman},
  {Steidel}, {Strom}, {Sugahara}, {Wilkins}, {Wofford}, \& {Xu}}]{berg22}
{Berg}, D.~A., {James}, B.~L., {King}, T., {et~al.} 2022, arXiv e-prints,
  arXiv:2203.07357.
\newblock \doarXiv{2203.07357}

\bibitem[{{Brinchmann}(2022)}]{brinchmann22}
{Brinchmann}, J. 2022, arXiv e-prints, arXiv:2208.07467.
\newblock \doarXiv{2208.07467}

\bibitem[{Cardelli {et~al.}(1989)Cardelli, Clayton, \& Mathis}]{cardelli89}
Cardelli, J.~A., Clayton, G.~C., \& Mathis, J.~S. 1989, \apj, 345, 245

\bibitem[{{Carnall} {et~al.}(2022){Carnall}, {Begley}, {McLeod}, {Hamadouche},
  {Donnan}, {McLure}, {Dunlop}, {Bondestam}, {Cullen}, {Jewell}, \&
  {Pollock}}]{carnall22}
{Carnall}, A.~C., {Begley}, R., {McLeod}, D.~J., {et~al.} 2022, arXiv e-prints,
  arXiv:2207.08778.
\newblock \doarXiv{2207.08778}

\bibitem[{{Chisholm} {et~al.}(2022){Chisholm}, {Saldana-Lopez}, {Flury},
  {Schaerer}, {Jaskot}, {et~al.}}]{chisholm22}
{Chisholm}, J., {Saldana-Lopez}, A., {Flury}, S., {et~al.} 2022, arXiv
  e-prints, arXiv:2207.05771.
\newblock \doarXiv{2207.05771}

\bibitem[{{Curti} {et~al.}(2017){Curti}, {Cresci}, {Mannucci}, {Marconi},
  {Maiolino}, \& {Esposito}}]{curti+17}
{Curti}, M., {Cresci}, G., {Mannucci}, F., {et~al.} 2017, \mnras, 465, 1384,
  \dodoi{10.1093/mnras/stw2766}

\bibitem[{{Curti} {et~al.}(2022){Curti}, {D'Eugenio}, {Carniani}, {Maiolino},
  {Sandles}, {Witstok}, {Baker}, {Bennett}, {Piotrowska}, {Tacchella},
  {Charlot}, {Nakajima}, {Maheson}, {Mannucci}, {Arribas}, {Belfiore},
  {Bonaventura}, {Bunker}, {Chevallard}, {Cresci}, {Curtis-Lake},
  {Hayden-Pawson}, {Kumari}, {Laseter}, {Looser}, {Marconi}, {Maseda}, {Jones},
  {Scholtz}, {Smit}, {Ubler}, \& {Wallace}}]{curti22a}
{Curti}, M., {D'Eugenio}, F., {Carniani}, S., {et~al.} 2022, arXiv e-prints,
  arXiv:2207.12375.
\newblock \doarXiv{2207.12375}

\bibitem[{{Dalcanton}(2007)}]{dalcanton07}
{Dalcanton}, J.~J. 2007, \apj, 658, 941

\bibitem[{{Dors} {et~al.}(2013){Dors}, {H{\"a}gele}, {Cardaci},
  {P{\'e}rez-Montero}, {Krabbe}, {V{\'\i}lchez}, {Sales}, {Riffel}, \&
  {Riffel}}]{Dors13}
{Dors}, O.~L., {H{\"a}gele}, G.~F., {Cardaci}, M.~V., {et~al.} 2013, \mnras,
  432, 2512, \dodoi{10.1093/mnras/stt610}

\bibitem[{{Draine}(2011)}]{draine11}
{Draine}, B.~T. 2011, {Physics of the Interstellar and Intergalactic Medium}

\bibitem[{{Ebeling} {et~al.}(2001){Ebeling}, {Jones}, {Fairley}, {Perlman},
  {Scharf}, \& {Horner}}]{ebeling01}
{Ebeling}, H., {Jones}, L.~R., {Fairley}, B.~W., {et~al.} 2001, \apjl, 548,
  L23, \dodoi{10.1086/318915}

\bibitem[{{Ferruit} {et~al.}(2022){Ferruit}, {Jakobsen}, {Giardino}, {Rawle},
  {Alves de Oliveira}, {et~al.}}]{ferruit22}
{Ferruit}, P., {Jakobsen}, P., {Giardino}, G., {et~al.} 2022, \aap, 661, A81,
  \dodoi{10.1051/0004-6361/202142673}

\bibitem[{{Finkelstein} {et~al.}(2019){Finkelstein}, {D'Aloisio},
  {Paardekooper}, {Ryan}, {Behroozi}, {et~al.}}]{finkelstein19}
{Finkelstein}, S.~L., {D'Aloisio}, A., {Paardekooper}, J.-P., {et~al.} 2019,
  \apj, 879, 36, \dodoi{10.3847/1538-4357/ab1ea8}

\bibitem[{{Finkelstein} {et~al.}(2022){Finkelstein}, {Bagley}, {Arrabal Haro},
  {Dickinson}, {Ferguson}, {Kartaltepe}, {Papovich}, {Burgarella}, {Kocevski},
  {Huertas-Company}, {Iyer}, {Larson}, {P{\'e}rez-Gonz{\'a}lez}, {Rose},
  {Tacchella}, {Wilkins}, {Chworowsky}, {Medrano}, {Morales}, {Somerville},
  {Yung}, {Fontana}, {Giavalisco}, {Grazian}, {Grogin}, {Kewley}, {Koekemoer},
  {Kirkpatrick}, {Kurczynski}, {Lotz}, {Pentericci}, {Pirzkal}, {Ravindranath},
  {Ryan}, {Trump}, {Yang}, {Almaini}, {Amor{\'\i}n}, {Annunziatella},
  {Backhaus}, {Barro}, {Behroozi}, {Bell}, {Bhatawdekar}, {Bisigello}, {Bromm},
  {Buat}, {Buitrago}, {Calabr{\'o}}, {Casey}, {Castellano}, {Ch{\'a}vez Ortiz},
  {Ciesla}, {Cleri}, {Cohen}, {Cole}, {Cooke}, {Cooper}, {Cooray}, {Costantin},
  {Cox}, {Croton}, {Daddi}, {Dav{\'e}}, {de la Vega}, {Dekel}, {Elbaz},
  {Estrada-Carpenter}, {Faber}, {Fern{\'a}ndez}, {Finkelstein}, {Freundlich},
  {Fujimoto}, {Garc{\'\i}a-Argum{\'a}nez}, {Gardner}, {Gawiser},
  {G{\'o}mez-Guijarro}, {Guo}, {Hamilton}, {Hathi}, {Holwerda}, {Hirschmann},
  {Hutchison}, {Jha}, {Jogee}, {Juneau}, {Jung}, {Kassin}, {Le Bail}, {Leung},
  {Lucas}, {Magnelli}, {Mantha}, {Matharu}, {McGrath}, {McIntosh}, {Merlin},
  {Mobasher}, {Newman}, {Nicholls}, {Pandya}, {Rafelski}, {Ronayne}, {Santini},
  {Seill{\'e}}, {Shah}, {Shen}, {Simons}, {Snyder}, {Stanway}, {Straughn},
  {Teplitz}, {Vanderhoof}, {Vega-Ferrero}, {Wang}, {Weiner}, {Willmer},
  {Wuyts}, \& {Zavala}}]{finkelstein22a}
{Finkelstein}, S.~L., {Bagley}, M.~B., {Arrabal Haro}, P., {et~al.} 2022, arXiv
  e-prints, arXiv:2207.12474.
\newblock \doarXiv{2207.12474}

\bibitem[{{Fontanot} {et~al.}(2014){Fontanot}, {Cristiani}, {Pfrommer},
  {Cupani}, \& {Vanzella}}]{fontanot14}
{Fontanot}, F., {Cristiani}, S., {Pfrommer}, C., {Cupani}, G., \& {Vanzella},
  E. 2014, \mnras, 438, 2097

\bibitem[{{Garnett}(1992)}]{garnett92}
{Garnett}, D.~R. 1992, \aj, 103, 1330

\bibitem[{{Garnett} {et~al.}(1995){Garnett}, {Skillman}, {Dufour},
  {et~al.}}]{garnett95}
{Garnett}, D.~R., {Skillman}, E.~D., {Dufour}, R.~J., {et~al.} 1995, \apj, 443,
  64

\bibitem[{Henry {et~al.}(2000)Henry, Edmunds, \& K\"oppen}]{henry00}
Henry, R. B.~C., Edmunds, M.~G., \& K\"oppen, J. 2000, \apj, 541, 660

\bibitem[{{Hutchison} {et~al.}(2019){Hutchison}, {Papovich}, {Finkelstein},
  {Dickinson}, {Jung}, {et~al.}}]{hutchison19}
{Hutchison}, T.~A., {Papovich}, C., {Finkelstein}, S.~L., {et~al.} 2019, \apj,
  879, 70, \dodoi{10.3847/1538-4357/ab22a2}

\bibitem[{{Izotov} {et~al.}(2016){Izotov}, {Schaerer}, {Thuan}, \&
  {others}}]{izotov16}
{Izotov}, Y.~I., {Schaerer}, D., {Thuan}, T.~X., \& {others}. 2016, \mnras,
  461, 3683

\bibitem[{{Izotov} {et~al.}(2012){Izotov}, {Thuan}, \& {Guseva}}]{izotov12}
{Izotov}, Y.~I., {Thuan}, T.~X., \& {Guseva}, N.~G. 2012, \aap, 546, A122

\bibitem[{{Jakobsen} {et~al.}(2022){Jakobsen}, {Ferruit}, {Alves de Oliveira},
  {Arribas}, {Bagnasco}, {Barho}, {Beck}, {Birkmann}, {B{\"o}ker}, {Bunker},
  {Charlot}, {de Jong}, {de Marchi}, {Ehrenwinkler}, {Falcolini}, {Fels},
  {Franx}, {Franz}, {Funke}, {Giardino}, {Gnata}, {Holota}, {Honnen}, {Jensen},
  {Jentsch}, {Johnson}, {Jollet}, {Karl}, {Kling}, {K{\"o}hler}, {Kolm},
  {Kumari}, {Lander}, {Lemke}, {L{\'o}pez-Caniego}, {L{\"u}tzgendorf},
  {Maiolino}, {Manjavacas}, {Marston}, {Maschmann}, {Maurer}, {Messerschmidt},
  {Moseley}, {Mosner}, {Mott}, {Muzerolle}, {Pirzkal}, {Pittet}, {Plitzke},
  {Posselt}, {Rapp}, {Rauscher}, {Rawle}, {Rix}, {R{\"o}del}, {Rumler},
  {Sabbi}, {Salvignol}, {Schmid}, {Sirianni}, {Smith}, {Strada}, {te Plate},
  {Valenti}, {Wettemann}, {Wiehe}, {Wiesmayer}, {Willott}, {Wright}, {Zeidler},
  \& {Zincke}}]{jakobsen22}
{Jakobsen}, P., {Ferruit}, P., {Alves de Oliveira}, C., {et~al.} 2022, \aap,
  661, A80, \dodoi{10.1051/0004-6361/202142663}

\bibitem[{{Kewley} {et~al.}(2019){Kewley}, {Nicholls}, \&
  {Sutherland}}]{kewley+19}
{Kewley}, L.~J., {Nicholls}, D.~C., \& {Sutherland}, R.~S. 2019, \araa, 57,
  511, \dodoi{10.1146/annurev-astro-081817-051832}

\bibitem[{{Lagache} {et~al.}(2018){Lagache}, {Cousin}, \&
  {Chatzikos}}]{lagache18}
{Lagache}, G., {Cousin}, M., \& {Chatzikos}, M. 2018, \aap, 609, A130,
  \dodoi{10.1051/0004-6361/201732019}

\bibitem[{{Li} {et~al.}(2019){Li}, {Narayanan}, \& {Dav{\'e}}}]{li19}
{Li}, Q., {Narayanan}, D., \& {Dav{\'e}}, R. 2019, \mnras, 490, 1425,
  \dodoi{10.1093/mnras/stz2684}

\bibitem[{{Luridiana} {et~al.}(2015){Luridiana}, {Morisset}, \&
  {Shaw}}]{luridiana15}
{Luridiana}, V., {Morisset}, C., \& {Shaw}, R.~A. 2015, \aap, 573, A42

\bibitem[{{Madau} \& {Haardt}(2015)}]{madau15}
{Madau}, P., \& {Haardt}, F. 2015, \apjl, 813, L8

\bibitem[{{Maiolino} \& {Mannucci}(2019)}]{maiolino+19}
{Maiolino}, R., \& {Mannucci}, F. 2019, \aapr, 27, 3,
  \dodoi{10.1007/s00159-018-0112-2}

\bibitem[{{M{\'e}ndez-Delgado} {et~al.}(2021){M{\'e}ndez-Delgado}, {Henney},
  {Esteban}, {Garc{\'\i}a-Rojas}, {Mesa-Delgado}, \&
  {Arellano-C{\'o}rdova}}]{mendez-delgado21}
{M{\'e}ndez-Delgado}, J.~E., {Henney}, W.~J., {Esteban}, C., {et~al.} 2021,
  \apj, 918, 27, \dodoi{10.3847/1538-4357/ac0cf5}

\bibitem[{{Naidu} {et~al.}(2020){Naidu}, {Tacchella}, {Mason}, {Bose}, {Oesch},
  \& {Conroy}}]{naidu20}
{Naidu}, R.~P., {Tacchella}, S., {Mason}, C.~A., {et~al.} 2020, \apj, 892, 109,
  \dodoi{10.3847/1538-4357/ab7cc9}

\bibitem[{{Nicholls} {et~al.}(2017){Nicholls}, {Sutherland}, {Dopita},
  {Kewley}, \& {Groves}}]{nicholls17}
{Nicholls}, D.~C., {Sutherland}, R.~S., {Dopita}, M.~A., {Kewley}, L.~J., \&
  {Groves}, B.~A. 2017, \mnras, 466, 4403

\bibitem[{{Peeples} \& {Shankar}(2011)}]{peeples11}
{Peeples}, M.~S., \& {Shankar}, F. 2011, \mnras, 417, 2962.
\newblock \doarXiv{1007.3743}

\bibitem[{{Piqu{\'e}ras} {et~al.}(2010){Piqu{\'e}ras}, {Legros}, {Pons},
  {Legay}, {Ferruit}, {Dorner}, {P{\'e}contal}, {Gnata}, \&
  {Mosner}}]{piqueras10}
{Piqu{\'e}ras}, L., {Legros}, E., {Pons}, A., {et~al.} 2010, in Society of
  Photo-Optical Instrumentation Engineers (SPIE) Conference Series, Vol. 7738,
  Modeling, Systems Engineering, and Project Management for Astronomy IV, ed.
  G.~Z. {Angeli} \& P.~{Dierickx}, 773812, \dodoi{10.1117/12.856860}

\bibitem[{{Pontoppidan} {et~al.}(2022){Pontoppidan}, {Blome}, {Braun}, {Brown},
  {Carruthers}, {Coe}, {DePasquale}, {Espinoza}, {Garcia Marin}, {Gordon},
  {Henry}, {Hustak}, {James}, {Koekemoer}, {LaMassa}, {Law}, {Lockwood},
  {Moro-Martin}, {Mullally}, {Pagan}, {Player}, {Proffitt}, {Pulliam},
  {Ramsay}, {Ravindranath}, {Reid}, {Robberto}, {Sabbi}, \&
  {Ubeda}}]{pontoppidan:22}
{Pontoppidan}, K., {Blome}, C., {Braun}, H., {et~al.} 2022, arXiv e-prints,
  arXiv:2207.13067.
\newblock \doarXiv{2207.13067}

\bibitem[{{Reddy} {et~al.}(2016){Reddy}, {Steidel}, {Pettini},
  {Bogosavljevi{\'c}}, \& {Shapley}}]{reddy16}
{Reddy}, N.~A., {Steidel}, C.~C., {Pettini}, M., {Bogosavljevi{\'c}}, M., \&
  {Shapley}, A.~E. 2016, \apj, 828, 108, \dodoi{10.3847/0004-637X/828/2/108}

\bibitem[{{Repp} \& {Ebeling}(2018)}]{repp18}
{Repp}, A., \& {Ebeling}, H. 2018, \mnras, 479, 844,
  \dodoi{10.1093/mnras/sty1489}

\bibitem[{{Rhoads} {et~al.}(2022){Rhoads}, {Wold}, {Harish}, {Kim}, {Pharo},
  {Malhotra}, {Gabrielpillai}, {Jiang}, \& {Yang}}]{rhoads22}
{Rhoads}, J.~E., {Wold}, I. G.~B., {Harish}, S., {et~al.} 2022, arXiv e-prints,
  arXiv:2207.13020.
\newblock \doarXiv{2207.13020}

\bibitem[{{Rodr{\'i}guez}(1999)}]{rodriguez99}
{Rodr{\'i}guez}, M. 1999, \aap, 348, 222.
\newblock \doarXiv{astro-ph/9906291}

\bibitem[{{Rodr{\'i}guez}(2002)}]{rodriguez02}
---. 2002, \aap, 389, 556, \dodoi{10.1051/0004-6361:20011823}

\bibitem[{{Rogers} {et~al.}(2022){Rogers}, {Skillman}, {Pogge}, {Berg},
  {Croxall}, {Bartlett}, {Arellano-C{\'o}rdova}, \& {Moustakas}}]{rogers22}
{Rogers}, N. S.~J., {Skillman}, E.~D., {Pogge}, R.~W., {et~al.} 2022, arXiv
  e-prints, arXiv:2209.03962.
\newblock \doarXiv{2209.03962}

\bibitem[{{Rogers} {et~al.}(2021){Rogers}, {Skillman}, {Pogge}, {Berg},
  {Moustakas}, {Croxall}, \& {Sun}}]{Rogers21}
---. 2021, \apj, 915, 21, \dodoi{10.3847/1538-4357/abf8b9}

\bibitem[{{Saldana-Lopez} {et~al.}(2022){Saldana-Lopez}, {Schaerer},
  {Chisholm}, {Flury}, {Jaskot}, {et~al.}}]{saldana-lopez22}
{Saldana-Lopez}, A., {Schaerer}, D., {Chisholm}, J., {et~al.} 2022, \aap, 663,
  A59, \dodoi{10.1051/0004-6361/202141864}

\bibitem[{{Sanders} {et~al.}(2022){Sanders}, {Jones}, \& {others}}]{sanders22}
{Sanders}, R.~L., {Jones}, T., \& {others}. 2022, \apj

\bibitem[{{Schaerer} {et~al.}(2022){Schaerer}, {Marques-Chaves}, {Oesch},
  {Naidu}, {Barrufet}, {Izotov}, {Guseva}, \& {Brammer}}]{schaerer22}
{Schaerer}, D., {Marques-Chaves}, R., {Oesch}, P., {et~al.} 2022, arXiv
  e-prints, arXiv:2207.10034.
\newblock \doarXiv{2207.10034}

\bibitem[{{Senchyna} {et~al.}(2019){Senchyna}, {Stark}, {Chevallard},
  {et~al.}}]{senchyna19}
{Senchyna}, P., {Stark}, D.~P., {Chevallard}, J., {et~al.} 2019, arXiv
  e-prints.
\newblock \doarXiv{1904.01615}

\bibitem[{{Senchyna} {et~al.}(2017){Senchyna}, {Stark}, {Vidal-Garc{\'{\i}}a},
  {et~al.}}]{senchyna17}
{Senchyna}, P., {Stark}, D.~P., {Vidal-Garc{\'{\i}}a}, A., {et~al.} 2017,
  \mnras, 472, 2608

\bibitem[{{Shapley} {et~al.}(2020){Shapley}, {Cullen}, {Dunlop}, {McLure},
  {Kriek}, {Reddy}, \& {Sanders}}]{shapley20}
{Shapley}, A.~E., {Cullen}, F., {Dunlop}, J.~S., {et~al.} 2020, \apjl, 903,
  L16, \dodoi{10.3847/2041-8213/abc006}

\bibitem[{{Shapley} {et~al.}(2019){Shapley}, {Sanders}, {Shao}, {Reddy},
  {Kriek}, {et~al.}}]{shapley19}
{Shapley}, A.~E., {Sanders}, R.~L., {Shao}, P., {et~al.} 2019, \apjl, 881, L35,
  \dodoi{10.3847/2041-8213/ab385a}

\bibitem[{{Stark}(2016)}]{stark16}
{Stark}, D.~P. 2016, \araa, 54, 761

\bibitem[{{Steidel} {et~al.}(2018){Steidel}, {Bogosavljevi{\'c}}, {Shapley},
  {et~al.}}]{steidel18}
{Steidel}, C.~C., {Bogosavljevi{\'c}}, M., {Shapley}, A.~E., {et~al.} 2018,
  \apj, 869, 123, \dodoi{10.3847/1538-4357/aaed28}

\bibitem[{{Steidel} {et~al.}(2016){Steidel}, {Strom}, {Pettini}, {Rudie},
  {Reddy}, \& {Trainor}}]{steidel16}
{Steidel}, C.~C., {Strom}, A.~L., {Pettini}, M., {et~al.} 2016, \apj, 826, 159

\bibitem[{{Tinsley}(1980)}]{tinsley80}
{Tinsley}, B.~M. 1980, \fcp, 5, 287

\bibitem[{{Topping} {et~al.}(2020){Topping}, {Shapley}, {Reddy}, {Sanders},
  {Coil}, {et~al.}}]{topping20}
{Topping}, M.~W., {Shapley}, A.~E., {Reddy}, N.~A., {et~al.} 2020, \mnras, 495,
  4430, \dodoi{10.1093/mnras/staa1410}

\bibitem[{Tremonti {et~al.}(2004)Tremonti, Heckman, Kauffmann,
  {et~al.}}]{tremonti04}
Tremonti, C.~A., Heckman, T.~M., Kauffmann, G., {et~al.} 2004, \apj, 613, 898

\bibitem[{{Trump} {et~al.}(2022){Trump}, {Arrabal Haro}, {et~al.}}]{trump22}
{Trump}, J.~R., {Arrabal Haro}, P., {et~al.} 2022, \apj

\bibitem[{{Wise} {et~al.}(2014){Wise}, {Demchenko}, {Halicek},
  {et~al.}}]{wise14}
{Wise}, J.~H., {Demchenko}, V.~G., {Halicek}, M.~T., {et~al.} 2014, \mnras,
  442, 2560

\bibitem[{{Zahid} {et~al.}(2014){Zahid}, {Dima}, {Kudritzki}, {Kewley},
  {Geller}, {Hwang}, {Silverman}, \& {Kashino}}]{zahid14}
{Zahid}, H.~J., {Dima}, G.~I., {Kudritzki}, R.-P., {et~al.} 2014, \apj, 791,
  130, \dodoi{10.1088/0004-637X/791/2/130}

\end{thebibliography}

\end{document}